\documentclass[aps,groupedaddress,nofootinbib,showkeys,showpacs,amsfonts,eqsecnum]{revtex4}
\usepackage{bm}

\newcommand{\tr}{{\textrm {tr}}}
\newcommand{\Tr}{{\textrm {Tr}}}

\newcommand{\la}{\left\langle}
\newcommand{\ra}{\right\rangle}

\begin{document}

\title{Covariant derivative expansion of the heat kernel}

\author{L.L. Salcedo}
\email{salcedo@ugr.es}

\affiliation{
Departamento de F\'{\i}sica Moderna, Universidad de Granada, E-18071
Granada, Spain }

\date{\today}

\begin{abstract}
Using the technique of labeled operators, compact explicit expressions
are given for all traced heat kernel coefficients containing zero,
two, four and six covariant derivatives, and for diagonal coefficients
with zero, two and four derivatives. The results apply to boundaryless flat
space-times and arbitrary non Abelian scalar and gauge background
fields.
\end{abstract}



\maketitle

\section{Introduction and conclusions}
\label{sec:1}

Ever since its introduction by Schwinger \cite{Schwinger:1951nm}, the
heat kernel of Laplace-type operators has become a useful tool to deal
with one loop effective actions in quantum field theory. This due to
the fact that the heat kernel provides a manifestly gauge invariant
regularization of ultraviolet divergences. An additional virtue, is
that, unlike the effective action, the heat kernel is a one valued
functional. The heat kernel can be applied to study spectral densities
of Klein-Gordon operators and in the proof of index theorems
\cite{Gilkey:1975iq,Atiyah:1973ad}, to compute the $\zeta$-function
\cite{Hawking:1977ja} and the anomalies of Dirac operators
\cite{Fujikawa:1980eg}, to deal with chiral gauge theories
\cite{Ball:1989xg} and models of QCD \cite{Bijnens:1996ww}, to the
Casimir effect \cite{Bordag:2001qi}, to compute black hole entropies
\cite{Callan:1994py}, etc. Exact calculations of the heat kernel at
coincident points are available in particular manifolds
\cite{Bytsenko:1996bc,Camporesi:1990wm} or for configurations
subjected to suitable algebraic constraints (of the constant curvature
type) \cite{Avramidi:1994zp,Avramidi:1995ik,Avramidi:1997jy}. In the
general case an asymptotic expansion in powers of the proper time, the
Seeley-DeWitt expansion \cite{Dewitt:1975ys,Seeley:1967ea}, is
available. The coefficients of the expansion have been computed to
rather high orders in several setups, including curved spaces with and
without boundary, and in presence of non Abelian gauge fields and non
Abelian scalar fields, using different methods
\cite{Ball:1989xg,Bel'kov:1996tn,vandeVen:1998pf,Moss:1999wq,%
Fliegner:1998rk,Avramidi:1991je,Gusynin:1989ky,Elizalde:1994bk,%
Vassilevich:2003xt}. The heat kernel expansion at finite temperature
has been discussed in \cite{Megias:2002vr,Megias:2003ui}. A
generalized heat kernel expansion around non c-number mass terms has
been introduced in
\cite{Osipov:2001bj1,Osipov:2001bj,Salcedo:2001qp}. The extension to
non commutative quantum field theory has been presented in
\cite{Vassilevich:2003yz}.

The standard heat kernel expansion can be regarded as a double expansion
in the strength and in the number of derivatives of the background
fields. It is therefore most suitable for external fields which are
both weak and adiabatic, i.e., of slow space-time variation. A
resummation of this expansion is provided by covariant perturbation
theory \cite{Barvinsky:1987uw}. Perturbation theory assumes weak but
not necessarily adiabatic fields. In this paper we study a different
resummation, namely, a covariant derivative expansion
\cite{Chan:1986jq,Salcedo:2000hp,Salcedo:2000hx}. The terms of this
expansion have a given number of covariant derivatives but any number
of scalar fields (these fields playing the role a non Abelian local
mass term). So the fields are assumed to be adiabatic (and the gauge
fields weak, to preserve gauge invariance) but the scalar fields may
be strong. Using the technique of labeled operators, we are able to
write in finite form the contributions to the traced heat kernel
classified by the number of derivatives. Such contributions can also
be regarded as generating functions for all the heat kernel
coefficients of the standard Seeley-DeWitt expansion with a fixed
number of derivatives. Explicit results are presented for zero, two,
four and six derivatives, involving 1, 1, 6 and 52 summands,
respectively. These results hold for boundaryless flat space-time but
non Abelian gauge and scalar fields. Extension to curved space-time
should also be possible using the symbols method. A covariant
derivative expansion to four derivatives has been obtained in
\cite{Gusynin:1990bu} for the diagonal heat kernel and in
\cite{Gusynin:1989nf} for the effective action, for curved space-time
in the case of minimal (i.e. Abelian) scalar field and no gauge
connection.

In Section \ref{sec:2} we define the covariant derivative expansion
for the trace of the heat kernel and work out the lower order terms.
Although the calculation could be done from scratch using the method
of symbols, a shorter path is provided by the method of Chan
\cite{Chan:1986jq} which was devised for the effective action. In that
Section results are presented to four derivatives. We show that after
expansion of our results in powers of the scalar field the standard
coefficients are recovered. Section \ref{sec:3} is devoted to
explaining the technique of labeled operators
\cite{Salcedo:1998sv,Garcia-Recio:2000gt,Salcedo:2000hp,Salcedo:2000hx}.
Such procedure allows to develop a calculus to deal with functions of
non commuting variables. The non commutative version of the ordinary
derivative is shown to satisfy the Leibniz and chain rules, as well as
to yield a non commutative version of the Taylor expansion. In Section
\ref{sec:4} the diagonal heat kernel coefficients, within the
derivative expansion, are obtained to four derivatives. This is done
through functional variation of the traced coefficients. Unlike the
standard expansion, in the resummed expansion one finds that each
diagonal term comes from the traced term of the same order. This
yields a consistency condition that is verified by our formulas.
Section \ref{sec:5} gives the six derivative contribution. It is based
on previous results for the effective action found in
\cite{Caro:1993fs}. Here one important issue is that of finding either
a short or a systematic expression for the result. This is due to the
existence of identities among the possible gauge invariant structures,
coming from integration by parts and Jacobi identities. In that
Section we give a relatively short expression for the six derivative
contribution, which contains 52 different gauge invariant
structures. In Section \ref{sec:6} we study the problem of finding a
standard basis of structures for the derivative expansion of generic
gauge invariant functionals. The corresponding problem in the context
of the standard heat kernel expansion has been treated before by
M\"uller \cite{Muller:1995bu,Muller:1996cq}. Some subtleties appear
for the derivative expansion because the analogous of the
Seeley-DeWitt coefficients are now functions (of labeled
operators). So, for instance, even if the elements of a basis are
complete and linearly independent it does not directly follow that the
coefficients must be unique; it is at least necessary to impose
permutation symmetry restrictions to the functions which play the role
of coefficients of the expansion. In this last Section we construct
standard basis of gauge invariant structures for functionals with
zero, two, four and six derivatives, with 1, 1, 6 and 37 elements,
respectively. It is worth noticing that, similarly to what happens for
the standard heat kernel expansion \cite{Vassilevich:2003yz}, all the
results presented here apply directly to non commutative quantum
field theory. This is particularly clear when such theories are
formulated within the quantum phase space approach
\cite{Alvarez-Gaume:2000dx}, which only requires to replace the
integral over coordinates by a trace on $X$-space. Indeed, at no place
do we use special commutation properties for our symbols, except
$[\partial_\mu,\partial_\nu]=0$ which holds in the non commutative
case too.

\section{Covariant derivative expansion of the heat kernel}
\label{sec:2}

Our goal is to obtain a derivative expansion for the heat kernel of
the Klein-Gordon operator
\begin{equation}
K= D_\mu^2+X \,.
\end{equation}
Here $X(x)$ is a multiplicative operator (i.e. an ordinary function)
which is a matrix in some internal space,
$D_\mu=\partial_\mu+V_\mu(x)$ is the covariant derivative, $V_\mu(x)$
being a matrix in internal space. Space-time is Euclidean and flat,
without boundaries and has dimension $d$. $K$ acts on matter fields in
the fundamental representation.

The standard heat kernel expansion is of the form
\begin{equation}
\langle x|e^{\tau K}|x\rangle=
\frac{1}{(4\pi\tau)^{d/2}}\sum_{n=0}^\infty \tau^n \,a_n(x)
\,.
\end{equation}
This is an asymptotic expansion where the $a_n$, known as (diagonal)
Seeley-DeWitt coefficients, are $\tau$-independent local gauge
covariant polynomials of dimension $2n$. They are constructed with $X$
and $D_\mu$ in gauge covariant combinations. One can choose to order
the expansion by powers of $\tau$ or, equivalently, by the mass
dimension carried by the external fields $V_\mu$ and $X$ and their
derivatives. The quantities $\partial_\mu$, $V_\mu$ and $X$ have
dimension 1,1,2 respectively. In spaces with boundary, $n$ may take
half-integer values \cite{McAvity:1991we}, but in our case the index
$n$ is a non negative integer.
The lowest order terms are
\begin{eqnarray}
a_0 &=& 1\,,\nonumber \\
a_1 &=& X\,, \nonumber \\
a_2 &=& \frac{1}{2}X^2+\frac{1}{6}X_{\mu\mu}+ \frac{1}{12}Z_{\mu\nu}^2 \,.
\end{eqnarray}

In the derivative expansion the terms are classified by the number of
covariant derivatives they carry, rather than the mass dimension they
carry (as in the standard heat kernel expansion) so
\begin{equation}
\langle x|e^{\tau K}|x\rangle =
\frac{1}{(4\pi\tau)^{d/2}} \sum_{n=0}^\infty  \tau^n A_n(x) \,,
\end{equation}
where $A_n(x)$ depends on $\tau$ and contains $2n$ covariant
derivatives. In this counting, $X$ counts as zeroth order, $[D_\mu,X]$
as first order, $Z_{\mu\nu}=[D_\mu,D_\nu]$ as second order, and so
on. (Counting the dimension carried by the background fields and by
$\tau$, $A_n$ has dimension $2n$.) Technically, the covariant
derivative expansion can be defined by introducing a bookkeeping
parameter $\lambda$ by means of $X(x)\to X(\lambda x)$ and
$V_\mu(x)\to \lambda V_\mu(\lambda x)$, and then expanding the
functional $\langle \lambda^{-1}x|e^{\tau
K(\lambda)}|\lambda^{-1}x\rangle$ in powers of $\lambda$. The
derivative expansion is a resummation of the standard expansion,
namely, if $a_n^q$ denotes the pieces of $a_n$ with exactly $2q$
covariant derivatives (and so with $n-q$ $X$'s)
\begin{equation}
A_q(x)= \sum_{n\ge q} \tau^{n-q}a_n^q(x) \,.
\end{equation}

In what follows we will set $\tau=1$, i.e. remove $\tau$ from the
formulas, since it can be restated at any moment by standard
dimensional counting. Thus, for instance
\begin{equation}
\langle x|e^K|x\rangle =
\frac{1}{(4\pi)^{d/2}} \sum_{n=0}^\infty  A_n(x) \,.
\end{equation}
In each $A_n$, $X$ appears to all orders. The prescription to restore
$\tau$ is simply to make the replacement $X\to\tau X$, plus $A_n \to
\tau^n A_n$.

For many purposes it is often sufficient to work with the functional
trace of the heat kernel,
\begin{equation}
\Tr\,e^K= \int d^dx\,\tr\,\langle x|e^K|x\rangle \,.
\end{equation}
The symbol $\Tr$ refers to the full trace on space-time and internal
spaces. We will use $\tr$ to denote the trace in the internal space
only. Introducing the short-hand notation\footnote{Our notational
conventions are summarized in Appendix \ref{app:A}.}
\begin{equation}
\la~~\ra := \frac{1}{(4\pi\tau)^{d/2}} \int
d^dx\,\tr\left(~~\right) \,
\end{equation}
the standard and derivative expansions take the form
\begin{eqnarray}
\Tr\, e^K &=&
\sum_{n=0}^\infty\la a_n(x) \ra
 = \sum_{n=0}^\infty  \la  A_n(x) \ra \,.
\end{eqnarray}

In $\Tr\, e^K$ one can use simpler coefficients, $b_n$ and $B_n$,
which coincide with $a_n$ and $A_n$, respectively, modulo by parts
integration and the trace cyclic property, so that 
\begin{equation}
\la b_n \ra = \la
a_n \ra \,,\quad \la B_n \ra = \la A_n \ra \,.
\label{eq:2.9}
\end{equation}
In particular,
\begin{eqnarray}
\Tr\, e^K &=& \sum_{n=0}^\infty  \la  B_n \ra \,.
\end{eqnarray}
Whereas the coefficients $A_n$ and the functionals $\la B_n\ra$ are
uniquely defined, there is an ambiguity in the choice of $B_n$ which
is exploited to choose them as simple as possible. The coefficients
$A_n$ can be obtained from the $B_n$ (see Section \ref{sec:4}).

The calculation of the coefficients $B_n$ can be done using the method
of symbols \cite{Salcedo:1996qy,Pletnev:1998yu} (actually, this
methods provides $A_n$ from which $B_n$ is obtained). In addition to
the derivation of the coefficients an important part of the
calculation is to find a simple expression for them, that is, removing
redundancies coming from the trace cyclic property, integration by
parts and Bianchi identities. Because much work has been devoted to
the covariant derivative expansion of the effective action, we have
found it practical to start from that functional where much of the
simplification work has already been done. The most useful results for
our purposes are found in the work of Chan \cite{Chan:1986jq} who
obtains the derivative expansion of the effective action with a
minimum of terms up to four derivatives. This work was later extended
to six derivatives in \cite{Caro:1993fs}. There we find (making
explicit terms up to second order)
\begin{equation}
\Tr\,\log K = \int\frac{d^dx d^dk}{(2\pi)^d}\tr\left(
-\log N+\frac{k^2}{d} N_\mu^2 + \cdots \right) \,,
\label{eq:2.7}
\end{equation}
where
\begin{equation}
N=(k^2-X)^{-1}\,,\quad N_\mu=[D_\mu,N] \,.
\end{equation}
In what follows we will consistently use the convention\footnote{Here
and elsewhere in this work $Y$ stands for a generic matrix-valued
function.} $Y_{\mu I}= [D_\mu,Y_I]$ where $Y_I$ is an object (such as
$N$, $X$ or $Z$) with an ordered set of Lorentz indices $I$. So for
instance\footnote{Note that in \cite{Caro:1993fs} the convention $Y_{
I\mu}= [D_\mu,Y_I]$ is used instead. Also $F_{\mu\nu}$ there
corresponds to $iZ_{\mu\nu}$ here.}
\begin{equation}
X_{\mu\nu}=[D_\mu,[D_\nu,X]] \,, 
\quad
Z_{\mu\nu\lambda}= [D_\mu,Z_{\nu\lambda}]=
[D_\mu,[D_\nu,D_\lambda]] \,.
\end{equation}

The formula (\ref{eq:2.7}) holds modulo a counter term action which
must be a local polynomial (in $X$, $V_\mu$ and $\partial_\mu$) of
degree at most $d$. Such counter terms depend on the renormalization
prescription chosen. The derivative expansion of the effective action
can be written as
\begin{equation}
\Tr\,\log K = \int\frac{d^dx d^dk}{(2\pi)^d}
\sum_{n=0}^\infty k^{2n}\frac{\Gamma(d/2)}{\Gamma(n+d/2)}
\tr\left( F_n \right) \,.
\label{eq:2.7a}
\end{equation}
Each $F_n$ is gauge covariant, contains a number $2n$ of $D_\mu$'s and
a number $2n$ of $N$'s, and has no explicit dependence on the
space-time dimension $d$. To four derivatives
\begin{eqnarray}
F_0 &=& -\log N \,, \nonumber \\ 
F_1 &=& \frac{1}{2}N_\mu^2 \,, \label{eq:2.11} \\
F_2 &=& N_\mu^2N_\nu^2
 - \frac{1}{2}(N_\mu N_\nu)^2 
 - (N N_{\mu\mu})^2
 - 2 N Z_{\mu\nu} NN_\mu N_\nu
 - \frac{1}{2} (Z_{\mu\nu}N^2)^2 \,.  \nonumber
\end{eqnarray}
These terms were obtained in \cite{Chan:1986jq}\footnote{Note that the
sign of the fourth term of $F_2$ is incorrect in
\cite{Chan:1986jq}.}. The six derivative term $F_3$ is given in
(\ref{eq:5.1}). It was obtained in \cite{Caro:1993fs} and contains 45
terms. In this Section we concentrate on terms up to four derivatives
and defer the treatment of the six derivative terms to Section
\ref{sec:5}. We will not consider terms with eight or more derivatives
in this work.

First of all we will translate the expansion (\ref{eq:2.7a}) to an
expansion for the heat kernel:
\begin{eqnarray}
e^K &=& \int_\Gamma \frac{dz}{2\pi i}\frac{e^z}{z-K}
= \int_\Gamma \frac{dz}{2\pi i}e^z\frac{d}{dz}\log(K-z)
= - \int_\Gamma \frac{dz}{2\pi i}e^z\log(K-z) \,.
\label{eq:2.14}
\end{eqnarray}
Here $\Gamma$ is a positively oriented path in the $z$ complex plane
enclosing the eigenvalues of $K$. Because the large eigenvalues of $K$
lie on the real negative axis, the path is taken starting and ending
at $-\infty$. Next we apply Chan's formula (\ref{eq:2.7a}) to
$\Tr\log(K-z)$, i.e., with the replacement $X \to X-z$. The first
thing to note is that the counter term ambiguities do not survive in
the heat kernel, since a polynomial in $z$ does not give a
contribution to the integral in (\ref{eq:2.14}). Taking $\Tr$ in
(\ref{eq:2.14}) and inserting (\ref{eq:2.7a}), one finds that all non
explicit dependence on $k_\mu$ comes in the form $X-k^2-z$. Making the
shift $z\to z-k^2$ removes all dependence on $k_\mu$ in $F_n$. (Such
change of variables is justified at the level of asymptotic
expansions we are considering.) Straightforward momentum integration
gives then
\begin{eqnarray}
\Tr\,e^K &=&
 -\sum_{n=0}^\infty
\int_\Gamma\frac{dz}{2\pi i}e^z 
\la F_n \ra \,,
\end{eqnarray}
that is,
\begin{eqnarray}
B_n &=& -\int_\Gamma\frac{dz}{2\pi i}e^z F_n
\label{eq:2.18}
\end{eqnarray}
(perhaps modulo integration by parts and cyclic property). In this
formula the quantities $F_n$ are given by the same expressions
(\ref{eq:2.11}) where now
\begin{equation}
N=(z-X)^{-1} \,.
\end{equation}

The integration over $z$ is easily done for the zeroth order term
(undoing the steps in (\ref{eq:2.14}))
\begin{eqnarray}
B_0 &=&  \int_\Gamma\frac{dz}{2\pi i}e^z \log N =
\int_\Gamma\frac{dz}{2\pi i}\frac{e^z}{z-X}  = e^X \,.
\end{eqnarray}

For $B_1$, we first expand the covariant derivative using the identity
\begin{equation}
N_\mu=NX_\mu N
\end{equation}
so that $z$ appears only in $N$ outside covariant derivatives
\begin{equation}
F_1 =\frac{1}{2}N^2X_\mu N^2X_\mu
\label{eq:2.22}
\end{equation}
(exploiting the cyclic property to move the last $N$ to the first
place).

When this $F_1$ is inserted in (\ref{eq:2.18}) the integral over $z$
cannot be readily done because $z$ appears in two places and the
operators do not commute in general. Here we apply the technique of
labeling the operators
\cite{Salcedo:1998sv,Garcia-Recio:2000gt,Salcedo:2000hp,Salcedo:2000hx}:
relative to the two $X_\mu$ in (\ref{eq:2.22}) there are three
positions, namely, the position 1 at the left of the first (leftmost)
$X_\mu$, the position 2, in between the two $X_\mu$ and the position
3, after the second $X_\mu$. (Operators at position 3 can be moved to
position 1 by the cyclic property.) The operators relative to which
the positions are defined (the two $X_\mu$ in this case) are named
``fixed operators''. The other operators are then labeled according to
their position relative to the fixed operators and moved to the left
(or to any convenient location in the expression). In this way we can
rewrite (\ref{eq:2.22}) as
\begin{equation}
F_1 =\frac{1}{2}N^2_1N^2_2X_\mu X_\mu= 
\frac{1}{2}\frac{1}{(z-X_1)^2}\frac{1}{(z-X_2)^2} X_\mu^2 \,.
\label{eq:2.22a}
\end{equation}
The labels 1 and 2 indicate at which position the labeled operators
$N_1$ and $N_2$ (or $X_1$, $X_2$) should be inserted (relative to the
fixed operators $X_\mu X_\mu$). The point of following this procedure
is that the labeled operators are effectively c-numbers: similar to
the time-ordered product or the normal product, they can be written in
any order since their true position is given by their label. Because
they are c-numbers, nothing prevents us from doing the $z$ integration
as for ordinary functions. We refer to Section \ref{sec:3} for
details on the use and properties of labeled operators. Their use is
crucial for the rest of the paper. After integration over $z$ we
obtain
\begin{equation}
\la B_1 \ra = \la f(X_1,X_2)X_\mu^2 \ra
\label{eq:2.24}
\end{equation}
where $X_1$ is $X$ located at position 1 and $X_2$ is $X$ at position
2. $f(X_1,X_2)$ is the ordinary function $f(x,y)$ evaluated at $X_1$
and $X_2$, with
\begin{eqnarray}
f(x,y) &=&  -\frac{1}{2}\int_\Gamma\frac{dz}{2\pi i}e^z 
\frac{1}{(z-x)^2}\frac{1}{(z-y)^2}  
= \frac{e^x-e^y}{(x-y)^3}-\frac{1}{2}\frac{e^x+e^y}{(x-y)^2} \,.
\end{eqnarray}
(By definition $\Gamma$ always encloses the poles, in this case at $x$
and $y$.)

The right hand side of (\ref{eq:2.24}) can be regarded as the
generating function for all the Seeley-DeWitt coefficients with two
covariant derivatives. Indeed, making a series expansion in $X$ one
obtains (using the cyclic property in the second equality)
\begin{eqnarray}
\la B_1 \ra &=& \la \left(
-\frac{1}{12}-\frac{1}{24}(X_1+X_2)
-\frac{1}{80}(X_1^2+X_2^2) -\frac{1}{60}X_1X_2
+\cdots \right)X_\mu^2 \ra
\nonumber \\
&=& \la 
-\frac{1}{12}X_\mu^2-\frac{1}{12}X X_\mu^2
-\frac{1}{40}X^2X_\mu^2 -\frac{1}{60}XX_\mu XX_\mu
+\cdots \ra \,.
\label{eq:2.24a}
\end{eqnarray}
To the order shown, this reproduces the pieces with two covariant
derivatives in $b_3$, $b_4$ and $b_5$ \cite{Bel'kov:1996tn}.

For the four derivative term we proceed similarly. Expanding the
covariant derivatives of $N$ one finds
\begin{eqnarray}
F_2 &=& 
 (N^2X_\mu N^2X_\mu)^2
-\frac{1}{2} (N^2X_\mu N^2X_\nu)^2
-4 (N^3 X_\mu N X_\mu)^2
- (N^3 X_{\mu\mu})^2
\nonumber \\ && 
-4 N^3X_\mu N X_\mu N^3 X_{\nu\nu}
-2 N^2 X_\mu N^2 X_\nu N^2 Z_{\mu\nu}
-\frac{1}{2} (N^2 Z_{\mu\nu})^2
\nonumber \\ &=&
 N_1^2 N_2^2 N_3^2 N_4^2 (X^2_\mu)^2
-\frac{1}{2} N_1^2 N_2^2N_3^2N_4^2(X_\mu X_\nu)^2
-4 N_1^3 N_2 N_3^3 N_4 (X_\mu^2)^2
- N_1^3 N_2^3 X_{\mu\mu}^2
\nonumber \\ && 
-4 N_1^3 N_2 N_3^3 X_\mu^2X_{\nu\nu}
-2 N_1^2  N_2^2 N_3^2 X_\mu X_\nu Z_{\mu\nu}
-\frac{1}{2} N_1^2 N_2^2 Z_{\mu\nu}^2 \,.
\label{eq:2.27}
\end{eqnarray}

The integrals over $z$ involved in the computation of the $B_n$ are of
the form
\begin{equation}
I_{r_1,r_2,\ldots,r_n}(X_1,X_2,\ldots,X_n) :=
\int_\Gamma\frac{dz}{2\pi i}e^z N_1^{r_1}N_2^{r_2}\cdots N_n^{r_n} \,.
\end{equation}
They can be computed from the basic integrals
\begin{equation}
I^0_n(X_1,X_2,\ldots,X_n):=\int_\Gamma\frac{dz}{2\pi i}e^z N_1N_2\cdots N_n
= \sum_{i=1}^ne^{X_i}\prod_{j\not= i}\frac{1}{X_i-X_j}
\end{equation}
using
\begin{equation}
I_{r_1,r_2,\ldots,r_n}(X_1,X_2,\ldots,X_n) =
\prod_{i=1}^n\frac{1}{(r_i-1)!}
\left(\frac{\partial}{\partial X_i}\right)^{r_i-1} 
I^0_n(X_1,X_2,\ldots,X_n) \,.
\end{equation}
(Alternatively, $I_{r_1,r_2,\ldots,r_n}(X_1,X_2,\ldots,X_n)$ can be
obtained from $I^0_{r_1+\cdots+r_n}$ taking the first $r_1$ arguments
to be $X_1$, then the next $r_2$ arguments to be $X_2$ and so on.)
Note that the functions $I_{r_1,r_2,\ldots,r_n}$ are everywhere
analytic in the $n$-dimensional complex plane, and invariant under a
common permutation of labels $r_j$ and arguments $X_j$. The basic
functions $I^0_n$ are completely symmetric and they can be obtained
from the following recurrence relation
\begin{eqnarray}
I^0_1(X_1) &=& e^{X_1} \,, \\
I^0_{n+1}(X_1,\ldots,X_{n-1},X_n,X_{n+1}) &=&
 \frac{I^0_n(X_1,\ldots,X_{n-1},X_n)-
I^0_n(X_1,\ldots,X_{n-1},X_{n+1})}{X_n-X_{n+1}} \,.
\nonumber
\end{eqnarray}

We will use the short-hand notation
\begin{equation}
I_{r_1,r_2,\ldots,r_n}:= I_{r_1,r_2,\ldots,r_n}(X_1,X_2,\ldots,X_n) \,.
\end{equation}
With this notation the lowest terms in the derivative expansion are
\begin{eqnarray}
\la B_0 \ra &=& \la I_1 \ra \,, \nonumber \\
\la B_1 \ra &=& \Big\langle -\frac{1}{2} I_{2,2} X_\mu^2 \Big\rangle \,, 
\label{eq:2.32} \\
\la B_2 \ra &=& \Big\langle
\left( -I_{2,2,2,2}+4I_{3,1,3,1}\right)(X_\mu^2)^2
+\frac{1}{2}I_{2,2,2,2}(X_\mu X_\nu)^2
+I_{3,3} X_{\mu\mu}^2
\nonumber \\ && ~~ 
+4 I_{3,1,3} X_\mu^2 X_{\nu\nu} 
+2I_{2,2,2}X_\mu X_\nu Z_{\mu\nu} + \frac{1}{2}I_{2,2}Z_{\mu\nu}^2
\Big\rangle \,.
\nonumber
\end{eqnarray}

Expanding in powers of $X$ one obtains for $B_2$
\begin{eqnarray}
\la B_2 \ra 
&=& \Big\langle
\frac{1}{12}Z_{\mu\nu}^2
\nonumber \\ && ~~
+\frac{1}{12}XZ_{\mu\nu}^2
\nonumber \\ && ~~
+\frac{1}{40}X^2Z_{\mu\nu}^2 + \frac{1}{60}(XZ_{\mu\nu})^2
+\frac{1}{60}X_\mu X_\nu Z_{\mu\nu}
+\frac{1}{120}X_{\mu\mu}^2
\label{eq:2.33} \\ && ~~
+\frac{1}{180}X^3Z_{\mu\nu}^2 + \frac{1}{120}X^2Z_{\mu\nu}X Z_{\mu\nu}
+\frac{1}{180}(XX_\mu X_\nu Z_{\mu\nu} + X_\mu XX_\nu Z_{\mu\nu}
X_\mu X_\nu XZ_{\mu\nu} )
+\frac{1}{180}X_\mu^2 X_{\nu\nu}
\nonumber \\ && ~~
+\cdots
\Big\rangle \,,
\nonumber
\end{eqnarray}
which reproduces $b_2$, $b_3$, $b_4$, $b_5$ to four covariant
derivatives \cite{Bel'kov:1996tn}.

The expressions (\ref{eq:2.32}), together with the similar expression
(\ref{eq:5.2}) for $B_3$, are the main result of this work. Note that,
compared to the standard heat kernel expansion, in the covariant
derivative expansion the numerical coefficients of the standard
expansion are replaced by coefficients which are functions of labeled
$X$'s.

It can be observed that the coefficient functions found are directly
consistent with the cyclic property. For instance, the identity
\begin{eqnarray}
\la f(X_1,X_2,X_3,X_4)X_\mu^2 X_\nu^2 \ra &=&
\la f(X_3,X_4,X_1,X_2)X_\nu^2 X_\mu^2 \ra
\end{eqnarray}
shows that one can always choose the coefficient function of
$(X_\mu^2)^2$ to be invariant under the cyclic permutation
$(X_1,X_2,X_3,X_4)\to (X_3,X_4,X_1,X_2)$ and this symmetry is explicit
in $\la B_2 \ra$ given in (\ref{eq:2.32}) using the permutation
symmetry properties of the functions $I_{r_1,r_2,\ldots,r_n}$.

There is another symmetry also realized in the $B_n$ which will play
an important role in what follows. This is {\em mirror} symmetry, that
is, the symmetry under transposition defined by linearity plus the
rules
\begin{equation}
(AB)^T=B^TA^T\,, \quad D_\mu^T=D_\mu\,,\quad X^T=X \,.
\end{equation}
They imply $[D_\mu,Y]^T=-[D_\mu,Y^T]$ and thus
\begin{equation}
X_{\mu_1\ldots\mu_n}^T= (-1)^nX_{\mu_1\ldots\mu_n} \,,\quad
Z_{\mu_1\ldots\mu_n}^T= (-1)^{n-1}Z_{\mu_1\ldots\mu_n} \,.
\end{equation}
In practice, because the total number of Lorentz indices is always
even, an equivalent rule is to pick up a minus sign for each
$Z_{\mu_1\ldots\mu_n}$ in the expression.

Mirror symmetry holds for the Klein-Gordon operator $K$ and for the
heat kernel $e^K$, and is manifest in $B_0$, $B_1$ and $B_2$. E.g.
\begin{equation}
\la f(X_1,X_2,X_3)X_\mu^2X_{\nu\nu} \ra^T
=
\la f(X_4,X_3,X_2)X_{\nu\nu}X_\mu^2 \ra
=
\la f(X_3,X_2,X_1)X_\mu^2 X_{\nu\nu}\ra
\end{equation}
shows that the coefficient function of $X_\mu^2X_{\nu\nu}$ (namely,
$I_{3,1,3}$ in this case) can be chosen even under transposition of
$X_1$ and $X_3$. It is curious that the coefficient function of $X_\mu
X_\nu Z_{\mu\nu}$ has a greater symmetry than required by cyclic and
mirror symmetries.

\section{Labeled operators}
\label{sec:3}

In this section we describe useful properties of labeled
operators. In an expression with labeled operators, there are {\em
fixed operators} relative to which the positions are defined and {\em
labeled operators} which carry position labels, e.g.\footnote{The
symbols $A$, $B$, $X$, etc, are generic and do not refer to those of
the heat kernel throughout this Section.}
\begin{equation}
f(A_1,B_2,C_3,\ldots)XY\cdots
\label{eq:B.1}
\end{equation}
$X$, $Y$, etc, are the fixed operators in this case, $A$ is to be
inserted before (to the left of) $X$ (position 1), $B$ between $X$ and
$Y$ (position 2), $C$ just after $Y$ (position 3) and so on. Such an
expression can be defined in two equivalent ways. First by writing $f$
as a sum of separable functions, e.g.
\begin{eqnarray}
f(x_1,x_2,x_3,\ldots) &=&
\sum_{n_1,n_2,n_3,\ldots}c_{n_1,n_2,n_3,\ldots}x_1^{n_1}x_2^{n_2}x_3^{n_3}\cdots \,,
\nonumber \\ 
f(A_1,B_2,C_3,\ldots)XY\cdots &:=&
\sum_{n_1,n_2,n_3,\ldots}c_{n_1,n_2,n_3,\ldots}A^{n_1}X B^{n_2}YC^{n_3}\cdots \,.
\end{eqnarray}
Another example would be the expansion of $f$ as a combination of plane
waves, through its Fourier transform.

Alternatively the expression in (\ref{eq:B.1}) can be interpreted
through its matrix elements. Taking $|n,A\rangle$ as a basis of
eigenvectors of $A$ with eigenvalue $a_n$, and similarly for $B$ and
$C$, and being $\langle n,A|$, etc, the corresponding dual basis,
\begin{eqnarray}
\langle n,A|f(A_1,B_2,C_3)XY|r,C\rangle &=&
\sum_m f(a_n,b_m,c_r) \langle n,A|X|m,B\rangle \langle m,B|Y|r,C\rangle \,.
\end{eqnarray}
The important point is that the expression in (\ref{eq:B.1}) depends
only on the function $f$ and not on how it is expanded.

The usefulness of the labeled operators stems from the fact that they
are effectively c-numbers since e.g. $A_1B_2=B_2A_1$, and so they can
be used in several ways. For instance, $[A,~]$ can be written as
$A_1-A_2$, since $[A,X]=(A_1-A_2)X$, then $f([A,~])$ can be
represented as $f(A_1-A_2)$. The well-known identity
\begin{equation}
e^{[A,~]}X= e^AXe^{-A}
\end{equation}
becomes trivial using labeled operators
\begin{equation}
e^{[A,~]}X= e^{A_1-A_2}X=e^{A_1}e^{-A_2}X=e^AXe^{-A} \,.
\end{equation}
This kind of properties have been used in
\cite{Garcia-Recio:2000gt,Salcedo:2000hp} to easily derive commutator
expansions.

Labeled operators appear naturally in non Abelian expansions. If one
needs to compute\footnote{By $f(A)$, and similar expressions, we mean
an ordinary function $f(x)$ evaluated at $x=A$, $A$ being an operator,
in the analytical extension sense, as in $e^A$ (and not to completely
general operator-valued functions of $A$). More generally, $f$ may be
matrix-valued taking values in a different space than that of $A$.}
$f(A+B)$ to first order in $B$, where $A$ and $B$ do not commute, a
standard technique is to transform this problem into $e^{A+B}$ by
means of a functional transform, and then apply Dyson's formula
\begin{equation}
e^{A+B}= e^A + \int_0^1 ds\, e^{sA}Be^{(1-s)A}+O(B^2) \,.
\end{equation}
Using labeled operators one can go further and obtain
\begin{equation}
e^{A+B}= e^A + \int_0^1 ds e^{sA_1+(1-s)A_2}B+O(B^2)
= e^A+ \frac{e^{A_1}-e^{A_2}}{A_1-A_2}B +O(B^2)
\end{equation}
($B$ being the fixed operator). Undoing now the functional transform
yields the useful relation, for a generic function $f(x)$,
\begin{equation}
f(A+B)= f(A) + \frac{f(A_1)-f(A_2)}{A_1-A_2}B +O(B^2) \,.
\end{equation}
This relation does not rely on the exponential function and can also
be established by using, e.g., a Taylor series expansion of
$f(x)$.

From the previous expansion we learn that under a {\em first order}
variation $\delta A$ of the operator $A$ (the function $f$ being
unchanged)
\begin{equation}
\delta\left( f(A) \right)=  \nabla f(A_1,A_2)\delta A
\label{eq:3.9}
\end{equation}
where the operation $\nabla$ is defined by
\begin{equation}
\nabla f(x_1,x_2):= \frac{f(x_1)-f(x_2)}{x_1-x_2} \,.
\label{eq:3.10}
\end{equation}
$\nabla$ maps a one-argument function $f(x)$ into a two-argument
function $\nabla f(x,y)$. Note that $\delta$ can be any variation of
$A$, including e.g. a derivative
\begin{equation}
\partial_\mu f(A)= \nabla f(A_1,A_2)\partial_\mu A \,
\end{equation}
or a commutator, $\delta=[X,~]$,
\begin{equation}
[X,f(A)]= \nabla f(A_1,A_2)[X,A] \,
\end{equation}
(indeed $\nabla f(A_1,A_2)[X,A]= \nabla f(A_1,A_2)(A_2-A_1)X=
(f(A_2)-f(A_1))X= [X,f(A)]$).

The operator $\nabla$ generalizes the ordinary derivative to the non
Abelian case. Of course, when $A$ and $\delta A$ commute the
right-hand side of (\ref{eq:3.9}) becomes $f^\prime(A)\delta A$
(applying L'H\^opital rule). It is straightforward to verify that the
operator $\nabla$ satisfies a Leibniz rule
\begin{equation}
\nabla (fg)(x_1,x_2)= \nabla f(x_1,x_2)g(x_2)+f(x_1)\nabla g(x_1,x_2) \,,
\end{equation}
where $f(x)$, $g(x)$ are possibly non commuting (matrix valued)
functions of a single variable. Furthermore, $\nabla$ also complies
with the chain rule
\begin{equation}
\nabla (f\circ g)(x_1,x_2)= \nabla f( g(x_1),g(x_2)) \nabla g
(x_1,x_2) \,
\end{equation}
(in this case $f(x)$ may be matrix valued) as is readily verified.

In the case of several variables one may need partial derivatives,
e.g.
\begin{equation}
\delta (f(A_1,B_2)X)=  \nabla_1 f(A_1,A_2,B_3)\delta A\, X
+\nabla_2 f(A_1,B_2,B_3)X\delta B 
+f(A_1,B_2)\delta X \,,
\end{equation}
where $\nabla_\ell$ indicates that it acts on the $\ell$-th argument
of $f$.

It is also convenient to define a $\nabla$ operator acting on the
space of functions of any number of variables, in such a way that if
maps a $n$-variable function $f(x_1,\ldots,x_n)$ to a $(n+1)$-variable
function, as
\begin{equation}
\nabla := \sum_k \nabla_k
\end{equation}
($\nabla_k$ acting on the $k$-th argument of the function) and so
\begin{eqnarray}
\nabla f(x_1,\ldots,x_{n+1}) &=&
\sum_{k=1}^n \frac{ f(x_1,\ldots,\widehat{x_{k+1}},\ldots,x_{n+1})
- f(x_1,\ldots,\widehat{x_{k}},\ldots,x_{n+1}) }{ x_k-x_{k+1} } 
 \,,
\end{eqnarray}
where $\widehat{x_\ell}$ indicates that the $\ell$-th argument is
missing. The definition is such that
\begin{equation}
\delta \left(f(A_1,\ldots,A_n)(\delta A)^{n-1} \right)
=
\nabla f(A_1,A_2,\ldots,A_{n+1}) (\delta A)^n
\end{equation}
for $\delta(A)=\delta A$ and $\delta(\delta A)=0$. Using this
operator we can write down the non Abelian version of Taylor's formula
\begin{equation}
f(A+B)= e^{B\nabla_A}f(A)
\end{equation}
($\nabla_A$ emphasizes that $\nabla$ acts on the $A$-dependence of the
expression). That is
\begin{eqnarray}
f(A+B) &=& f(A)+\nabla f(A_1,A_2) B
+\frac{1}{2!}\nabla^2 f(A_1,A_2,A_3) B^2+\cdots
\nonumber \\
&=&
f(A)+
\frac{f(A_1)-f(A_2)}{A_1-A_2}B 
\nonumber \\ &&
+
\left(\frac{f(A_1)}{(A_2-A_1)(A_3-A_1)}
+\frac{f(A_2)}{(A_1-A_2)(A_3-A_2)}
+\frac{f(A_3)}{(A_1-A_3)(A_2-A_3)}
\right) B^2 +\cdots   \,.
\label{eq:3.20}
\end{eqnarray}
In the Abelian case the coefficients reduce to those of the standard
Taylor expansion.

To finish this Section we note a very important point when using
labeled operators, namely, that of the regularity of the functions of
the type $f(A_1,A_2,...)$ at the coincidence limit of two or more
arguments. These functions must be regular (free from poles) at the
coincidence limit to define meaningful operators. For instance, an
expression of the type
\begin{equation}
\frac{1}{A_1-A_2}B
\end{equation}
is only formal as it refers to any solution $Y$ of the equation
$[A,Y]=B$. Depending on the operators $A$ and $B$ such solution either
does not exist or is not unique. The operation $\nabla$ (cf. eq.
(\ref{eq:3.10})) does not introduce singularities (poles) in that
limit and the functions appearing in the expansion (\ref{eq:3.20}) are
all regular.

\section{Diagonal heat kernel coefficients}
\label{sec:4}

The diagonal coefficients $A_n(x)$ can be computed using e.g. the
method of symbols, however, having the $B_n$ it is simpler to derive
them from the relation \cite{Ball:1989xg}
\begin{equation}
\delta_X \Tr \,e^{D_\mu^2+X}
= \Tr \left(e^{D_\mu^2+X}\delta X \right)
\end{equation}
where $\delta_X$ is a first order variation with respect to $X$. That is
\begin{equation}
\langle x|e^K|x\rangle= \frac{\delta \Tr \,e^K}{\delta X(x)} \,.
\end{equation}
For the Seeley-DeWitt coefficients this implies
\begin{equation}
a_n(x)= (4\pi)^{d/2}\frac{\delta \la b_{n+1}\ra }{\delta X(x)} \,,
\end{equation}
whereas for the derivative expansion coefficients gives
\begin{equation}
A_n(x)= (4\pi)^{d/2}\frac{\delta \la B_n\ra }{\delta X(x)} \,.
\label{eq:3.4}
\end{equation}
This relation allows to obtain $A_n$ from $B_n$, but $B_n$ can also be
obtained from $A_n$ though (\ref{eq:2.9}). This implies the
consistency condition
\begin{equation}
\la B_n \ra= (4\pi)^{d/2} \la \frac{\delta \la B_n\ra }{\delta X(x)} \ra 
\,.
\end{equation}
It can be shown that this consistency condition is satisfied by our
expressions (\ref{eq:2.32}). This only requires the property
\begin{equation}
I_{r_1,r_2,\ldots,r_n} = e^{X_1} f(X_2-X_1,\ldots,X_n-X_1)
\end{equation}
(this relation codifies the heat kernel property $e^K \to e^a e^K$
under the shift $X\to X+a$, $a$ being a c-number). The consistency
condition is not sufficient to determine $B_n$ since it will be
satisfied too by all heat kernel-like operators of the form
$\exp(X+f(D_\mu^2))$.

Using the results in Section \ref{sec:3} relative to the manipulation of
labeled operators, one can carry out the functional derivative with
respect to $X$ indicated in (\ref{eq:3.4}) for $\la B_n\ra$ given in
(\ref{eq:2.32}). However, in the present case it is simpler to go back
to $F_n$ (i.e., before integration over $z$ and labeling of operators)
and do the variation there, using the identity
\begin{equation}
\delta_X N= N\delta X N \,.
\end{equation}
We illustrate the method with $A_1$:
\begin{eqnarray}
\delta_X\la F_1 \ra &=&
\delta_X\la \frac{1}{2}N^2X_\mu N^2X_\mu \ra =
\la 
N\delta X N^2X_\mu N^2X_\mu
+N^2\delta X N X_\mu N^2 X_\mu
+ N^2 (\delta X)_\mu N^2X_\mu
\ra
\nonumber \\ &=&
\la 
\left( N^2X_\mu N^2X_\mu N
+ N X_\mu N^2 X_\mu N^2
-   (N^2X_\mu N^2)_\mu
\right) \delta X 
\ra
\label{eq:3.8}
 \\ &=&
\big\langle
 \big(
- 2 N^2 X_\mu N X_\mu N^2 
- N^2 X_{\mu\mu} N^2
\big) \delta X
\big\rangle \,.
\nonumber
\end{eqnarray}
I.e.
\begin{eqnarray}
\frac{\delta\la F_1 \ra}{\delta X} &=&
-\frac{1}{(4\pi)^{d/2}}
\left(
 2 N_1^2 N_2 N^2_3 X_\mu  X_\mu 
+ N^2_1 N^2_2 X_{\mu\mu}
\right) \,.
\end{eqnarray}

Following the same procedure with $F_2$ and carrying out the
integration over $z$, we find
\begin{eqnarray}
A_0 &=& I_1 \,, \nonumber \\
A_1 &=& 2 I_{2,1,2} X_\mu^2 + I_{2,2} X_{\mu\mu} \,, \nonumber \\
A_2 &=&
  2 I_{2, 1, 2} \, Z_{\mu \nu} Z_{\mu \nu} 
\nonumber \\ && 
+ ( 4 I_{2, 2, 2} - 8 I_{3, 0, 3} + 8 I_{2, 1, 3} )\, Z_{\mu \mu \nu} X_{\nu} 
\nonumber \\ && 
+   (4 I_{2, 1, 2, 2} - 16 I_{3, 0, 1, 3})\, 
     X_{\mu} Z_{\mu \nu} X_{\nu} 
\nonumber \\ && 
+ (32 I_{3, 0, 1, 3} - 16 I_{2, 1, 1, 3} - 8 I_{2, 1, 2, 2} + 
      16 I_{3, 0, 2, 2})\,  Z_{\mu \nu} X_{\mu} X_{\nu} 
\nonumber \\ && 
 + 2 I_{3, 3} \, X_{\mu \mu \nu \nu} 
\nonumber \\ && 
+ ( 2 I_{2, 3, 2} + 4 I_{3, 1, 3} + 4 I_{2, 2, 3} )\, 
     X_{\mu \mu}  X_{\nu \nu} 
\nonumber \\  &&
+ 8 I_{3, 1, 3} \, X_{\mu \nu} X_{\mu \nu} 
\nonumber \\  &&
+  (16 I_{3, 1, 3} + 8 I_{3, 2, 2})\,  X_{\mu \nu \nu} X_{\mu} 
\nonumber \\ && 
+ (  4 I_{2, 2, 2, 2} +  16 I_{3, 1, 1, 3} + 16 I_{2, 2, 1, 3})\,  
X_{\mu} X_{\mu \nu} X_{\nu} 
\nonumber \\ && 
+ (  2 I_{2, 2, 2, 2} + 8 I_{3, 1, 1, 3}
   + 8 I_{2, 2, 1, 3} )\,  X_{\mu} X_{\nu \nu} X_{\mu} 
\nonumber \\ && 
+ (  4 I_{2, 2, 2, 2} + 16 I_{3, 1, 1, 3} + 8 I_{2, 2, 1, 3} 
+ 8 I_{2, 3, 1, 2} + 8 I_{3, 1, 2, 2} + 8 I_{3, 2, 1, 2})\, 
     X_{\mu \mu} X_{\nu} X_{\nu} 
\nonumber \\ && 
+ (32 I_{3, 1, 1, 3} + 16 I_{3, 1, 2, 2})\, X_{\mu \nu} X_{\mu} X_{\nu} 
\nonumber \\ && 
  + (4 I_{2, 2, 1, 2, 2}  +  16 I_{3, 1, 1, 1, 3} 
 + 16 I_{2, 1, 2, 1, 3} + 8 I_{2, 2, 2, 1, 2} + 8 I_{2, 1, 3, 1, 2} +
      16 I_{2, 2, 1, 1, 3} 
)\,  X_{\mu} X_{\mu} X_{\nu} X_{\nu} 
\nonumber \\ && 
 + (  4 I_{2, 2, 1, 2, 2} + 16 I_{3, 1, 1, 1, 3} + 16 I_{2, 2, 1, 1, 3}
)\,  X_{\mu} X_{\nu} X_{\mu} X_{\nu} 
\nonumber \\ && 
 + ( 4 I_{2, 2, 1, 2, 2} + 16 I_{3, 1, 1, 1, 3} + 16 I_{2, 2, 1, 1, 3}
)\,  X_{\mu} X_{\nu} X_{\nu} X_{\mu} \,.
\end{eqnarray}

$A_1$ is a direct transcription of (\ref{eq:3.8}). However, in $A_2$
we have shortened the expression by using a standard convention,
namely, we have identified every term with its mirror conjugate and
have used only one of the two forms. In other words, each term stands
for the semi sum of itself plus its mirror conjugate. For instance,
\begin{eqnarray}
4 I_{2, 2, 3} \, X_{\mu \mu}  X_{\nu \nu} 
&:=& (2 I_{2, 2, 3}+ 2 I_{3, 2, 2} )\, X_{\mu \mu}  X_{\nu \nu} \,,
\nonumber \\
 16 I_{3, 0, 2, 2}\, Z_{\mu \nu} X_{\mu} X_{\nu} 
&:=&
8 I_{3, 0, 2, 2}\, Z_{\mu \nu} X_{\mu} X_{\nu}  
+8 I_{2, 2, 0, 3}\,  X_{\mu} X_{\nu} Z_{\mu \nu} \,.
\label{eq:4.11}
\end{eqnarray}

In obtaining $A_2$ we have used the Jacobi identity (here $Y$
represents an arbitrary quantity)
\begin{equation}
Y_{\mu\nu}=Y_{\nu\mu}+[Z_{\mu\nu},Y]
\label{eq:3.11} 
\end{equation}
to reduce the number of terms, e.g. by eliminating terms of the type
$X_{\mu\nu}Z_{\mu\nu}$ (as $\frac{1}{2}[Z_{\mu\nu},X]Z_{\mu\nu}$) and
by canonically ordering the Lorentz indices.

\section{Six derivative terms, $B_3$}
\label{sec:5}

The term $F_3$ in the expansion (\ref{eq:2.7a}) has been computed in
\cite{Caro:1993fs}:


\begin{eqnarray}
F_3&=&
+\frac{20}{3}N_{\alpha}N_{\alpha}N_{\beta}N_{\beta}N_{\gamma}N_{\gamma}
-2N_{\alpha}N_{\beta}N_{\alpha}N_{\gamma}N_{\beta}N_{\gamma}
+\frac{2}{3}N_{\alpha}N_{\beta}N_{\gamma}N_{\alpha}N_{\beta}N_{\gamma}
+2N_{\alpha}N_{\alpha}N_{\beta}N_{\gamma}N_{\gamma}N_{\beta}
\nonumber\\&&
-4N_{\alpha}N_{\alpha}N_{\beta}N_{\gamma}N_{\beta}N_{\gamma}
-16NN_{\alpha}N_{\beta}N_{\alpha}N_{\gamma}N_{\beta\gamma}
+16NN_{\alpha}N_{\beta}N_{\beta}N_{\gamma}N_{\alpha\gamma}
+6N^2N_{\alpha\alpha}N_{\beta}N_{\beta}N_{\gamma\gamma}
\nonumber\\&&
-16NN_{\alpha\alpha}NN_{\beta\beta}N_{\gamma}N_{\gamma}
-8NN_{\alpha}N_{\alpha\beta}NN_{\gamma}N_{\gamma\beta}
-16NN_{\alpha}N_{\alpha\beta}N_{\beta}NN_{\gamma\gamma}
-8NN_{\alpha}N_{\beta}NN_{\alpha\gamma}N_{\beta\gamma}
\nonumber\\&&
+8NN_{\alpha}N_{\beta}NN_{\beta\gamma}N_{\alpha\gamma}
+10NN_{\alpha}N_{\beta\beta}NN_{\alpha}N_{\gamma\gamma}
-4NN_{\alpha}N_{\beta\beta}NN_{\gamma\gamma}N_{\alpha}
+8NN_{\alpha}N_{\beta\beta}N_{\alpha}NN_{\gamma\gamma}
\nonumber\\&&
-\frac{8}{3}NN_{\alpha\alpha}NN_{\beta\beta}NN_{\gamma\gamma}
+12N^2N_{\alpha\alpha}N_{\beta}NN_{\beta\gamma\gamma}
+3N^2N_{\alpha\beta\beta}N^2N_{\alpha\gamma\gamma}
\nonumber\\&&
-8Z_{\alpha\beta}N_{\gamma}NN_{\alpha}N_{\beta}NN_{\gamma}
+4NZ_{\alpha\beta}NN_{\gamma}N_{\alpha}N_{\beta}N_{\gamma}
-8NZ_{\alpha\beta}NN_{\alpha}N_{\beta}N_{\gamma}N_{\gamma}
-8NZ_{\alpha\beta}NN_{\alpha}N_{\gamma}N_{\beta}N_{\gamma}
\nonumber\\&&
-4NZ_{\alpha\beta}NN_{\alpha}N_{\gamma}N_{\gamma}N_{\beta}
-16NZ_{\alpha\beta}NN_{\alpha}NN_{\gamma\gamma}N_{\beta}
-16Z_{\alpha\beta}N^2N_{\gamma\gamma}NN_{\alpha}N_{\beta}
-16Z_{\alpha\beta}N^2N_{\gamma}N_{\gamma\alpha}NN_{\beta}
\nonumber\\&&
-16NZ_{\alpha\beta}N^2N_{\alpha\gamma}N_{\gamma}N_{\beta}
-16NZ_{\alpha\beta}N^2N_{\alpha}N_{\gamma\gamma}N_{\beta}
-8N^2Z_{\alpha\beta}N^2N_{\alpha\gamma}N_{\beta\gamma}
-16N^2Z_{\alpha\beta}N^2N_{\alpha}N_{\beta\gamma\gamma}
\nonumber\\&&
-16NZ_{\alpha\beta}N^2N_{\alpha\gamma\gamma}NN_{\beta}
+8N^2Z_{\alpha\alpha\beta}N^2N_{\beta}N_{\gamma\gamma}
\nonumber\\&&
-2N^3Z_{\alpha\alpha\beta}N^3Z_{\gamma\beta\gamma}
+16Z_{\alpha\beta}N^3Z_{\gamma\alpha\gamma}N^2N_{\beta}
+NZ_{\alpha\beta}NN_{\gamma}NZ_{\alpha\beta}NN_{\gamma}
-4NZ_{\alpha\beta}NN_{\alpha}NZ_{\beta\gamma}NN_{\gamma}
\nonumber\\&&
-4N^2Z_{\alpha\beta}N^2N_{\gamma}Z_{\alpha\beta}N_{\gamma}
+8Z_{\alpha\beta}N^2N_{\gamma}Z_{\alpha\gamma}N^2N_{\beta}
+16Z_{\alpha\beta}N^3Z_{\alpha\gamma}N_{\gamma}NN_{\beta}
-2NZ_{\alpha\beta}N^2Z_{\alpha\beta}NN_{\gamma}N_{\gamma}
\nonumber\\&&
-4NZ_{\alpha\beta}N^2Z_{\alpha\gamma}NN_{\beta}N_{\gamma}
+4NZ_{\alpha\beta}N^2Z_{\alpha\gamma}NN_{\gamma}N_{\beta}
-8Z_{\alpha\beta}N^3Z_{\alpha\beta}N^2N_{\gamma\gamma}
\nonumber\\&&
-\frac{4}{3}N^2Z_{\alpha\beta}N^2Z_{\alpha\gamma}N^2Z_{\beta\gamma}
\,.
\label{eq:5.1}
\end{eqnarray}


In this formula the number of explicit terms has been reduced by
identifying terms related by i) cyclic permutations and ii) mirror
symmetry. As noted before some conventions here differ from those in
\cite{Caro:1993fs}. In order to obtain the heat kernel coefficient
$B_3$ from $F_3$ we use the procedure of Section \ref{sec:2}, as in
e.g. (\ref{eq:2.27}). This yields


\begin{eqnarray}
\langle B_3 \rangle &=& \Big\langle
 -3 I_{4  4}X_{\alpha \beta \beta} X_{\alpha \gamma \gamma}
 -24 I_{4  1  4}X_{\alpha} X_{\beta \alpha} X_{\beta \gamma \gamma}
+(-12 I_{3  2  4} - 12 I_{4  1  4}) 
 X_{\alpha} X_{\beta \beta} X_{\alpha \gamma \gamma} 
 \nonumber\\&& 
 + \frac{8}{3} I_{3  3  3}X_{\alpha \alpha} X_{\beta \beta} X_{\gamma \gamma}
+ (-24 I_{4  1  1  4} - 24 I_{4  1  2  3}) 
 X_{\alpha} X_{\alpha} X_{\beta} X_{\beta \gamma \gamma}
 -12 I_{4  1  1  4}X_{\alpha} X_{\beta} X_{\alpha} X_{\beta \gamma \gamma}
\nonumber\\&& 
+ (-12 I_{1  3  2  4} - 6 I_{1  4  1  4} - 6 I_{2  2  2  4} + 
 16 I_{2  2  3  3} + 4 I_{2  3  2  3} + 16 I_{3  1  3  3}) 
 X_{\alpha} X_{\alpha} X_{\beta \beta} X_{\gamma \gamma}
\nonumber\\&&
+ (-24 I_{1  4  1  4} - 24 I_{2  3  1  4}) 
 X_{\alpha} X_{\beta} X_{\alpha \beta} X_{\gamma \gamma}
 +8 I_{3  2  3  2} 
X_{\alpha} X_{\beta} X_{\alpha \gamma} X_{\beta \gamma}
 -8 I_{3  2  3  2} 
 X_{\alpha} X_{\beta} X_{\beta \gamma} X_{\alpha \gamma}
\nonumber\\&& 
 -24 I_{1  4  1  4}X_{\alpha} X_{\beta} X_{\gamma \beta} X_{\gamma \alpha}
+ (-24 I_{1  4  1  4} - 24 I_{2  3  1  4} + 
 16 I_{3  2  2  3})X_{\alpha} X_{\alpha \beta} X_{\beta} X_{\gamma \gamma} 
+ 8 I_{3  2  3  2}X_{\alpha} X_{\alpha \beta} X_{\gamma} X_{\gamma \beta} 
 \nonumber\\&&
 -24 I_{4  1  4  1} 
 X_{\alpha} X_{\beta \alpha} X_{\gamma} X_{\beta \gamma}
 +(-6 I_{1  4  1  4} - 12 I_{1  4  2  3} - 10 I_{2  3  2  3} - 
 8 I_{2  3  3  2})X_{\alpha} X_{\beta \beta} X_{\alpha} X_{\gamma \gamma} 
 \nonumber\\&&
 +(-24 I_{1  3  2  1  4} - 24 I_{1  4  1  1  4} - 
 24 I_{2  2  2  1  4} + 32 I_{2  2  3  1  3} - 24 I_{2  3  1  1  4} + 
 16 I_{2  3  2  1  3} + 32 I_{3  1  2  2  3} + 32 I_{3  1  3  1  3} 
 \nonumber\\&&\qquad
+  16 I_{3  2  1  2  3})
X_{\alpha} X_{\alpha} X_{\beta} X_{\beta} X_{\gamma \gamma}
\nonumber\\&&
+ (-48 I_{4  1  1  4  1} - 48 I_{4  1  2  3  1}) 
 X_{\alpha} X_{\alpha} X_{\beta} X_{\gamma} X_{\beta \gamma} 
 +16 I_{3  2  1  3  2}
X_{\alpha} X_{\alpha} X_{\beta} X_{\gamma} X_{\gamma \beta}
\nonumber\\&&
 +(-16 I_{1  3  2  3  2} - 16 I_{2  2  2  3  2} + 
 32 I_{3  1  3  2  2} - 48 I_{4  1  1  4  1} - 48 I_{4  1  2  3  1}) 
 X_{\alpha} X_{\alpha} X_{\beta} X_{\beta \gamma} X_{\gamma}
\nonumber\\&& 
+ (-24 I_{1  1  3  2  4} - 24 I_{1  1  4  1  4} - 16 I_{2  1  2  3  3} - 
 40 I_{2  1  3  2  3} - 24 I_{2  1  4  1  3} - 16 I_{3  1  3  2  2}) 
 X_{\alpha} X_{\alpha} X_{\beta} X_{\gamma \gamma} X_{\beta}
\nonumber\\&& 
 +(-24 I_{1  4  1  1  4} - 24 I_{2  3  1  1  4} + 16 I_{3  2  1  2  3}) 
 X_{\alpha} X_{\beta} X_{\alpha} X_{\beta} X_{\gamma \gamma}
\nonumber\\&& 
+ (16 I_{3  2  2  2  2} + 16 I_{3  2  3  1  2} - 48 I_{4  1  1  4  1}) 
 X_{\alpha} X_{\beta} X_{\alpha} X_{\gamma} X_{\beta \gamma} 
\nonumber\\&& 
 +(-16 I_{2  1  3  2  3} + 16 I_{3  2  1  3  2}) 
 X_{\alpha} X_{\beta} X_{\alpha} X_{\gamma} X_{\gamma \beta}
+ 16 I_{3  2  3  1  2}
X_{\alpha} X_{\beta} X_{\gamma} X_{\alpha} X_{\beta \gamma}
\nonumber\\&&
 +(-24 I_{1  1  4  1  1  4} - 48 I_{1  1  4  1  2  3} - 
 8 I_{1  2  1  3  2  3} - 16 I_{1  2  2  2  2  3} + 
 32 I_{1  2  3  1  3  2} + 16 I_{2  1  3  1  2  3} + 
 64 I_{2  1  3  1  3  2}
\nonumber\\&&\qquad
 - 24 I_{2  1  4  1  2  2} - 
 \frac{20}{3} I_{2  2  2  2  2  2} + \frac{64}{3} I_{3  1  3  1  3  1}) 
 X_{\alpha} X_{\alpha} X_{\beta} X_{\beta} X_{\gamma} X_{\gamma} 
 \nonumber\\&& 
+ (-48 I_{1  1  4  1  1  4} + 8 I_{1  2  1  3  2  3} + 
 16 I_{1  2  2  2  2  3} + 16 I_{1  2  3  1  2  3} - 
 16 I_{1  3  2  3  1  2} - 48 I_{2  1  4  1  1  3} - 
 16 I_{2  2  2  2  1  3} 
\nonumber\\&&\qquad
+ 4 I_{2  2  2  2  2  2} + 
 32 I_{3  1  3  2  1  2})X_{\alpha} X_{\alpha} X_{\beta} X_{\gamma} 
X_{\beta} X_{\gamma}
\nonumber\\&&
+ (-48 I_{1  1  3  2  1  4} - 
 24 I_{1  1  4  1  1  4} + 8 I_{1  2  3  1  2  3} - 
 32 I_{2  1  2  3  1  3} - 40 I_{2  1  3  2  1  3} - 
 2 I_{2  2  2  2  2  2})X_{\alpha} X_{\alpha} X_{\beta} X_{\gamma} 
X_{\gamma} X_{\beta}
\nonumber\\&&
+ (16 I_{1  2  1  3  2  3} - 
 8 I_{2  1  3  2  3  1} + 16 I_{2  2  1  3  2  2} + 
 2 I_{2  2  2  2  2  2} + 8 I_{3  1  2  3  1  2} - 
 12 I_{4  1  1  4  1  1})X_{\alpha} X_{\beta} X_{\alpha} X_{\gamma} 
X_{\beta} X_{\gamma}
\nonumber\\&&
 +(8 I_{1  2  1  3  2  3} - \frac{2}{3} I_{2  2  2  2  2  2})
X_{\alpha} X_{\beta} X_{\gamma} X_{\alpha} X_{\beta}X_{\gamma}
\nonumber
\end{eqnarray}
\begin{eqnarray}
 && + 8 I_{3  2  3}X_{\alpha} X_{\beta \beta} Z_{\gamma \alpha \gamma} 
 +(16 I_{2  3  3} + 16 I_{3  2  3}) 
X_{\alpha} X_{\beta \gamma \gamma} Z_{\alpha \beta}
+8 I_{3  2  3}X_{\alpha \beta} X_{\gamma \beta} Z_{\alpha \gamma}
 \nonumber\\&& 
 -16 I_{3  1  2  3}X_{\alpha} X_{\alpha} X_{\beta} Z_{\gamma \beta \gamma}
+ 16 I_{3  1  2  3} 
 X_{\alpha} X_{\beta} X_{\gamma \alpha} Z_{\beta \gamma}
 \nonumber\\&&
+ 16 I_{3  2  3  1}X_{\alpha} X_{\alpha \beta} X_{\gamma} Z_{\beta \gamma}
+ (16 I_{2  2  2  3} + 32 I_{2  3  1  3} + 16 I_{3  1  2  3} + 
 32 I_{3  2  1  3})X_{\alpha} X_{\beta} X_{\gamma \beta} 
Z_{\alpha \gamma}
\nonumber\\&&
+ (16 I_{1  2  3  3} + 16 I_{2  3  1  3} + 
 16 I_{3  2  1  3})X_{\alpha} X_{\beta} X_{\gamma \gamma} 
Z_{\alpha \beta} 
+ (32 I_{3  1  2  3} + 32 I_{3  1  3  2}) 
 X_{\alpha} X_{\beta \alpha} X_{\gamma} Z_{\beta \gamma}
\nonumber\\&& 
 +(16 I_{2  2  2  3} + 16 I_{2  2  3  2} + 16 I_{2  3  1  3} + 
 16 I_{3  1  2  3})X_{\alpha} X_{\beta \beta} X_{\gamma} 
Z_{\alpha \gamma}
\nonumber\\&&
 +(8 I_{2  2  2  2  2} + 32 I_{3  1  1  2  3} + 
 32 I_{3  1  1  3  2} + 32 I_{3  1  3  2  1} + 16 I_{3  2  1  3  1}) 
 X_{\alpha} X_{\alpha} X_{\beta} X_{\gamma} Z_{\beta \gamma} 
 \nonumber\\&& 
+ (8 I_{2  2  2  2  2} + 32 I_{3  1  1  2  3} + 32 I_{3  1  1  3  2} + 
 16 I_{3  1  2  1  3} + 16 I_{3  1  2  2  2} + 16 I_{3  2  1  3  1}) 
 X_{\alpha} X_{\beta} X_{\alpha} X_{\gamma} Z_{\beta \gamma}
\nonumber\\&& 
+ (32 I_{2  2  1  2  3} + 32 I_{2  2  1  3  2} + 16 I_{2  2  2  1  3} + 
 4 I_{2  2  2  2  2} + 32 I_{2  3  1  1  3} + 32 I_{3  1  1  2  3} + 
 8 I_{3  1  2  1  3})X_{\alpha} X_{\beta} X_{\beta} X_{\gamma} 
Z_{\alpha \gamma} 
 \nonumber\\&&
+ (8 I_{1  3  2  3  1} - 4 I_{2  2  2  2  2} + 
 8 I_{3  1  2  1  3})X_{\alpha} X_{\beta} X_{\gamma} X_{\alpha} Z_{\beta \gamma}
\nonumber
\end{eqnarray}
\begin{eqnarray}
&& + 2 I_{3  3}Z_{\alpha \alpha \beta} Z_{\gamma \beta \gamma} 
 + 16 I_{3  1  3}X_{\alpha} Z_{\alpha \beta} Z_{\gamma \beta \gamma}
+ 8 I_{1  3  3}X_{\alpha \alpha} Z_{\beta \gamma} Z_{\beta \gamma} 
 +(16 I_{1  1  3  3} + 2 I_{2  2  2  2})X_{\alpha} X_{\alpha} 
Z_{\beta \gamma} Z_{\beta \gamma}
\nonumber\\&&
+ 4 I_{2  2  2  2} 
 X_{\alpha} X_{\beta} Z_{\alpha \gamma} Z_{\beta \gamma}
 +(-16 I_{1  3  1  3} - 4 I_{2  2  2  2}) 
 X_{\alpha} X_{\beta} Z_{\beta \gamma} Z_{\alpha \gamma}
+ (4 I_{2  2  2  2} + 8 I_{3  1  3  1})X_{\alpha} Z_{\alpha \beta} 
X_{\gamma} Z_{\beta \gamma}
\nonumber\\&&
 +(4 I_{1  3  3  1} - I_{2  2  2  2}) 
 X_{\alpha} Z_{\beta \gamma} X_{\alpha} Z_{\beta \gamma}
\nonumber \\&&
\nonumber \\
 && + \frac{4}{3} I_{2  2  2}
 Z_{\alpha \beta} Z_{\alpha \gamma} Z_{\beta \gamma}
\Big\rangle
\,.
\label{eq:5.2}
\end{eqnarray}


For the proper interpretation of this formula, it is important to
recall that mirror symmetric terms have been identified, as in
(\ref{eq:4.11}).

$F_3$ contains 45 terms but only 40 different structures of fixed
operators, since some of the terms differ only by the position of
underivated $N$'s. Similarly, $B_3$ contains 52 different {\em
structures} of fixed operators (e.g. $ X_{\alpha} Z_{\beta \gamma}
X_{\alpha} Z_{\beta \gamma}$), each with a coefficient function
written as a combination of functions
$I_{r_1,r_2,\ldots,r_n}$. Counting each of these as different, $B_3$
contains a total of 147 {\em terms}.

As explained at length in \cite{Caro:1993fs}, the expression of $F_3$
is not unique due to integration by parts and the Jacobi identity
(\ref{eq:3.11}) and this is also true for $B_3$. The previous
expression for $B_3$ comes directly from $F_3$ using only cyclic and
mirror symmetries to reduce the number of terms. This is also the
shortest expression we have found for $B_3$ from the point of view of
the number of terms. No systematic minimization of the number of terms
in $B_3$ has been attempted (it was done in \cite{Caro:1993fs} for
$F_3$), nevertheless the existence of a much shorter expression seems
unlikely.  Alternatively, one can try to reduce the number of
structures. The length defined from the point of view of the number of
structures can be reduced from 52 to 37. This is because, as discussed
below, all functionals of the type of $B_3$ can be written using a
standard basis of structures with 37 elements. (Conceivably, a
concrete functional such as $B_3$ could be written using a smaller
number of structures, but this is unlikely.) The price to pay for a
smaller number of structures is to increase the number of terms from
147 to about two thousand terms.

\section{Basis of structures with two, four and six derivatives}
\label{sec:6}

Instead of finding shortest expressions, it is also of interest to
find a standard basis \cite{Muller:1995bu,Muller:1996cq} of
structures. We discuss that problem in this Section.

Specifically, one would like to express a generic gauge
invariant functional $F(D,X)$ constructed with $D_\mu$ and $X$ and
with a fixed number of covariant derivatives, as a combination of
structures $T_i$ (the basis, independent of the functional $F$) with
$F$-dependent coefficients $F^{(i)}$ which are functions of labeled
$X$'s:
\begin{equation}
\la F \ra = \sum_{i=1}^N \la F^{(i)}(X_1,X_2,\ldots)T_i \ra \,
\label{eq:6.1}
\end{equation}
(the number of arguments in $F^{(i)}$ being the number of fixed
operators in the structure $T_i$).

A standard basis is subject to some requirements, namely, i) the
structures in the basis must be sufficient to express any functional
and ii) all of them must be necessary (i.e., no one can be removed
from the basis without spoiling the requirement (i)).

The expressions for $B_0$, $B_1$ and $B_2$ in (\ref{eq:2.32}) suggest
that the following are standard basis for gauge invariant functionals
with cyclic and mirror symmetry: For zero derivatives,
\begin{eqnarray}
T_1= 1 \,.
\end{eqnarray}
For functionals with two derivatives
\begin{eqnarray}
T_1= X_\mu^2 \,,
\end{eqnarray}
and for four covariant derivatives,
\begin{eqnarray}
T_1= (X_\mu^2)^2 \,,\quad
T_2=(X_\mu X_\nu)^2  \,,\quad
T_3=  X_{\mu\mu}^2  \,,\quad
T_4=  X_\mu^2 X_{\nu\nu}  \,,\quad
T_5= X_\mu X_\nu Z_{\mu\nu}  \,,\quad
T_6= Z_{\mu\nu}^2 \,.
\end{eqnarray}
This is indeed so. For instance, for two covariant derivatives, in
addition to $T_1= X_\mu^2$, one could write down a further structure
$X_{\mu\mu}$, however, this is redundant since
\begin{equation}
\la f(X)\,X_{\mu\mu} \ra 
=\la -[D_\mu, f(X)]\,X_\mu \ra
=\la -\nabla f(X_1,X_2)\,X_\mu^2 \ra
=\la F^{(1)}(X_1,X_2)\,T_1 \ra \,.
\end{equation}

As we have noted at the end of Section \ref{sec:3} the coefficient
functions $F^{(i)}(X_1,X_2,...)$ must be regular in the coincidence
limit of two or more arguments to define meaningful expressions. If
this requirement is not attended one finds that formally a smaller
number of structures would suffice. For four derivatives these are
$T_1$ and $T_2$. For instance, $T_3$ can be reduced to $T_1$ as
follows
\begin{eqnarray}
\la f(X_1,X_2) X_{\mu\mu}^2 \ra &=& 
  \la f(X_1,X_3)
(X_1-2X_2+X_3)(X_3-2X_4+X_1)D_\mu D_\mu D_\nu D_\nu \ra 
\nonumber \\
&=&  \la f(X_1,X_3)
\frac{(X_1-2X_2+X_3)(X_3-2X_4+X_1)}{(X_2-X_1)(X_3-X_2)(X_4-X_3)(X_1-X_4)}
X_\mu X_\mu X_\nu X_\nu \ra
\nonumber \\ 
&=& \la F^{(1)}(X_1,X_2,X_3,X_4)\,T_1 \ra \,.
\label{eq:6.6}
\end{eqnarray}
Such reduction is faulty as the identity used in the second step
$D_\mu\to (X_2-X_1)^{-1}X_\mu$ is only formal (as noted above, the
equation $[Y,X]=X_\mu$ does not imply $Y=D_\mu$).

In addition to being sufficient and necessary, for a standard basis
one may ask whether the coefficients $F^{(i)}$ are unique. In general
they will not be unique. For instance, for the structure $X_\mu^2$, to
any given $F^{(1)}(X_1,X_2)$, one can add an arbitrary odd function
$f(X_1,X_2)=-f(X_2,X_1)$. Such an addition $\la f(X_1,X_2)X_\mu^2\ra$
vanishes identically using cyclic symmetry. Therefore,
$F^{(1)}(X_1,X_2)$ can only be unique if one imposes the further
condition that it must be symmetric under transposition of $X_1$ and
$X_2$. It is clear that under such restriction $F^{(1)}(X_1,X_2)$ is
unique. (This can be verified using the technique of bare structures
introduced below.) Equivalently, a functional $\la
F^{(1)}(X_1,X_2)X_\mu^2\ra$ is identically zero if and only if the
symmetric function $F^{(1)}(X_1,X_2)$ is identically zero.

In the four derivative case, the coefficients can always be chosen to
have the following symmetries
\begin{eqnarray}
&& F^{(1)}_{1234}=F^{(1)}_{1432}=F^{(1)}_{3214}=F^{(1)}_{3412} \,,
\nonumber\\
&& F^{(2)}_{1234}=F^{(2)}_{1432}=F^{(2)}_{2143}=F^{(2)}_{2341}
=F^{(2)}_{3214}=F^{(2)}_{3412}=F^{(2)}_{4123}=F^{(2)}_{4321} \,,
\nonumber\\
&& F^{(3)}_{12}=F^{(3)}_{21} \,,
\nonumber\\ 
&& F^{(4)}_{123}=F^{(4)}_{321}
\nonumber\\
&& F^{(5)}_{123}=F^{(5)}_{321} \,,
\nonumber\\
&& F^{(6)}_{12}=F^{(6)}_{21} \,,
\end{eqnarray}
where we use a shorthand notation e.g.
$F^{(2)}_{1432}=F^{(2)}(X_1,X_4,X_3,X_2)$. For instance
\begin{equation}
\la  F^{(5)}_{123}X_\mu X_\nu Z_{\mu\nu} \ra
= -\la  F^{(5)}_{432} Z_{\mu\nu}X_\nu X_\mu \ra
= -\la  F^{(5)}_{321} X_\nu X_\mu Z_{\mu\nu}\ra
= \la  F^{(5)}_{321} X_\mu X_\nu Z_{\mu\nu}\ra
\end{equation}
(using, in the first equality, that a term and its mirror symmetric
have been identified). With these symmetry restrictions, these
$F^{(i)}$ can be shown to be unique.

We finally come to the six derivative case (always assuming gauge
invariant functionals with cyclic and mirror symmetries). The 52
structures appearing for $B_3$ in (\ref{eq:5.2}) are neither necessary
nor sufficient. A standard basis is as follows
\begin{eqnarray}
&& T_{1} = X_{\alpha \beta \gamma} X_{\alpha \beta \gamma} 
\,,\quad
 T_{2} =X_{\alpha} X_{\beta\gamma} X_{\alpha \beta \gamma} 
\,,\quad
T_{3} = X_{\alpha \beta} X_{\alpha \gamma} X_{\beta \gamma}  
\,,\quad
T_{4}=X_{\alpha} X_{\alpha} X_{\beta \gamma} X_{\beta \gamma}
  \,, \quad 
\nonumber\\&&
T_{5} = X_{\alpha} X_\beta X_\gamma X_{\alpha \beta \gamma}  
  \,, \quad 
T_{6} = X_{\alpha} X_{\beta} X_{\alpha \gamma} X_{\beta \gamma}
  \,, \quad 
T_{7} = X_{\alpha} X_{\beta} X_{\beta \gamma} X_{\alpha \gamma}
  \,, \quad 
T_{8} = X_{\alpha} X_{\alpha \beta} X_{\gamma} X_{\gamma \beta}
  \,, \quad 
\nonumber\\&&
T_{9} = X_{\alpha} X_{\beta \gamma} X_{\alpha} X_{\beta \gamma}
  \,, \quad 
T_{10} = X_{\alpha} X_{\alpha} X_{\beta} X_{\gamma} X_{\beta \gamma}
  \,, \quad 
T_{11} = X_{\alpha} X_{\beta} X_{\beta} X_{\gamma} X_{\alpha \gamma}
  \,, \quad 
T_{12} = X_{\alpha} X_{\beta} X_{\alpha} X_{\gamma} X_{\beta \gamma}
\,, \quad 
\nonumber\\&&
T_{13} = X_{\alpha} X_{\beta} X_{\gamma} X_{\alpha} X_{\beta \gamma}
\,, \quad 
T_{14} = X_{\alpha} X_{\alpha} X_{\beta} X_{\beta} X_{\gamma} X_{\gamma}
\,, \quad 
T_{15} = X_{\alpha} X_{\alpha} X_{\beta} X_{\gamma} X_{\beta} X_{\gamma}
\,, \quad 
\nonumber\\&&
T_{16} = X_{\alpha} X_{\alpha} X_{\beta} X_{\gamma} X_{\gamma} X_{\beta}
\,, \quad 
T_{17} = X_{\alpha} X_{\beta} X_{\alpha} X_{\gamma} X_{\beta} X_{\gamma}
\,, \quad 
T_{18} = X_{\alpha} X_{\beta} X_{\gamma} X_{\alpha} X_{\beta} X_{\gamma}
\,, \quad 
\nonumber\\&&
T_{19} = X_{\alpha} X_{\beta \gamma} Z_{\beta\alpha \gamma}
\,, \quad 
T_{20} = X_{\alpha \beta} X_{\alpha \gamma} Z_{\beta \gamma}
\,, \quad 
T_{21} = X_{\alpha} X_{\beta} X_{\gamma} Z_{\alpha \beta \gamma}
\,, \quad 
T_{22} = X_{\alpha} X_{\beta} X_{\alpha \gamma} Z_{\beta \gamma}
\,, \quad 
\nonumber\\&&
T_{23} = X_{\alpha} X_{\beta} X_{\beta \gamma} Z_{\alpha \gamma}
\,, \quad 
T_{24} = X_{\alpha} X_{\alpha \beta} X_{\gamma} Z_{\beta \gamma}
\,, \quad 
T_{25} = X_{\alpha} X_{\alpha} X_{\beta} X_{\gamma} Z_{\beta \gamma}
\,, \quad 
T_{26} = X_{\alpha} X_{\beta} X_{\alpha} X_{\gamma} Z_{\beta \gamma}
\,, \quad 
\nonumber\\&&
T_{27} = X_{\alpha} X_{\beta} X_{\beta} X_{\gamma} Z_{\alpha \gamma}
\,, \quad 
T_{28} = X_{\alpha} X_{\beta} X_{\gamma} X_{\alpha} Z_{\beta \gamma}
\,, \quad 
\nonumber\\&&
T_{29} = Z_{\alpha \beta \gamma} Z_{\alpha \beta \gamma}
\,, \quad 
T_{30} = X_{\alpha} Z_{\beta\gamma} Z_{\alpha \beta \gamma}
\,, \quad 
T_{31} = X_{\alpha \beta} Z_{\alpha \gamma} Z_{\beta \gamma}
\,, \quad 
T_{32} = X_{\alpha} X_{\alpha} Z_{\beta \gamma} Z_{\beta \gamma}
\,, \quad 
\nonumber\\&&
T_{33} = X_{\alpha} X_{\beta} Z_{\alpha \gamma} Z_{\beta \gamma}
\,, \quad 
T_{34} = X_{\alpha} X_{\beta} Z_{\beta \gamma} Z_{\alpha \gamma}
\,, \quad 
T_{35} = X_{\alpha} Z_{\alpha \beta} X_{\gamma} Z_{\beta \gamma}
\,, \quad 
T_{36} = X_{\alpha} Z_{\beta \gamma} X_{\alpha} Z_{\beta \gamma}
\,, \quad 
\nonumber\\&&
T_{37} = Z_{\alpha \beta} Z_{\alpha \gamma} Z_{\beta \gamma}
\,.
\label{eq:5.11}
\end{eqnarray}

To establish that this set is sufficient we follow the ideas put
forward by M\"uller in \cite{Muller:1996cq} for the standard heat
kernel expansion. Consider the set of all possible structures with six
derivatives, removing those that are redundant using i) dummy indices,
ii) cyclic symmetry and iii) mirror symmetry. There are 211 such
distinct structures.  Using integration by parts one can always remove
all structures where Lorentz indices are contracted within the same
factor, e.g. the index $\alpha$ in $ X_{\alpha\alpha} X_{\beta \gamma}
X_{\beta \gamma}$. Next, one can use the Jacobi identity
(\ref{eq:3.11}) to choose the order of the covariant derivatives
within each factor, for instance, if $X_{\alpha \beta \gamma}
X_{\alpha \beta \gamma}$ is retained, $ X_{\alpha \beta \gamma}
X_{\alpha \gamma \beta} $ becomes redundant. For the same reason all
structures of the type $(\cdots
Y_{\cdots\alpha\beta\cdots\gamma}\cdots Z_{\cdots\alpha\beta})$ are
also redundant. Finally, 
$X_{\alpha} X_{\beta} X_{\gamma} Z_{\beta\alpha \gamma}$
can be reduced to $T_{21}$  using the Bianchi identity
\begin{equation}
Z_{\alpha\beta\gamma}=Z_{\beta\alpha\gamma}+Z_{\gamma\beta\alpha} \,.
\end{equation}
This produces the set of structures in (\ref{eq:5.11}).

The above constructive procedure shows that the 37 structures are
sufficient. Before showing that they are also necessary, we need to
discuss the symmetries of their coefficient functions $F^{(i)}$. These
are as follows
\begin{eqnarray}
%
&&
F^{(1)}_{12}=F^{(1)}_{21} 
,\quad  
F^{(3)}_{123}=F^{(3)}_{213}=F^{(3)}_{321}
,\quad 
F^{(4)}_{1234}=F^{(4)}_{3214}
,\quad 
F^{(5)}_{1234}=F^{(5)}_{4321}
,
\nonumber \\ &&
F^{(6)}_{1234}=F^{(6)}_{3214}
,\quad 
F^{(7)}_{1234}=F^{(7)}_{3214}
,\quad 
F^{(8)}_{1234}=F^{(8)}_{3412}
,\quad 
F^{(9)}_{1234}=F^{(9)}_{2143}=F^{(9)}_{3412}
,
\nonumber \\ &&
F^{(11)}_{12345}=F^{(11)}_{32154}
,\quad 
F^{(13)}_{12345}=F^{(13)}_{54321}
,\quad 
F^{(14)}_{123456}=F^{(14)}_{165432}=F^{(14)}_{321654}
,
\nonumber \\ &&
F^{(15)}_{123456}=F^{(15)}_{321654}
,\quad 
F^{(16)}_{123456}=F^{(16)}_{321654}=F^{(16)}_{456123}
,\quad 
F^{(17)}_{123456}=F^{(17)}_{165432}=F^{(17)}_{456123}
,
\nonumber \\ &&
F^{(18)}_{123456}=F^{(18)}_{654321}=F^{(18)}_{234561}
,\quad 
F^{(20)}_{123}=F^{(20)}_{321}
,\quad 
F^{(27)}_{12345}=F^{(27)}_{54321}
,
\nonumber \\ &&
F^{(28)}_{12345}=F^{(28)}_{54321}
,\quad 
F^{(29)}_{12}=F^{(29)}_{21}
,\quad 
F^{(31)}_{123}=F^{(31)}_{213}
,\quad 
F^{(32)}_{1234}=F^{(32)}_{3214}
,
\nonumber \\ &&
F^{(33)}_{1234}=F^{(33)}_{3214}
,\quad 
F^{(34)}_{1234}=F^{(34)}_{3214}
,\quad 
F^{(35)}_{1234}=F^{(35)}_{3412}
,\quad 
F^{(36)}_{1234}=F^{(36)}_{4321}=F^{(36)}_{3412}
,
\nonumber \\ &&
F^{(37)}_{123}=F^{(37)}_{213}=F^{(37)}_{231} 
\,.
\end{eqnarray}
We have indicated only the generators of the symmetry group (e.g. for
$F^{(3)}$ it follows that this function is completely symmetric under
permutations). As discussed before, only after imposing the
symmetries can one expect the coefficient functions to be unique for a
given functional. A subtlety that did not appear in the two or four
derivative cases is that one has to take into account not only true
symmetries but also {\em quasi-symmetries}. For instance,
\begin{eqnarray}
&&\la  F^{(31)}_{123} T_{31} \ra=
\la  F^{(31)}_{123} X_{\alpha \beta} Z_{\alpha \gamma} Z_{\beta \gamma} \ra
= \la  F^{(31)}_{432} Z_{\beta \gamma}Z_{\alpha \gamma} X_{\alpha \beta} \ra
= \la  F^{(31)}_{213} X_{\alpha \beta}Z_{\beta \gamma}Z_{\alpha \gamma} \ra
\nonumber \\ &&
= \la  F^{(31)}_{213} X_{\beta \alpha}Z_{\alpha \gamma}Z_{\beta \gamma} \ra
= \la  F^{(31)}_{213} ( X_{\alpha\beta}Z_{\alpha \gamma}Z_{\beta \gamma}
+ [Z_{\beta \alpha},X]Z_{\alpha \gamma}Z_{\beta \gamma}) \ra
\nonumber \\ &&
= \la  F^{(31)}_{213} X_{\alpha\beta}Z_{\alpha \gamma}Z_{\beta \gamma} 
+  (X_1-X_2)F^{(31)}_{213} 
Z_{\alpha\beta}Z_{\alpha \gamma}Z_{\beta \gamma} \ra
\nonumber \\ &&
= \la  F^{(31)}_{213} T_{31} +  (X_1-X_2)F^{(31)}_{213} T_{37}  \ra \,,
\end{eqnarray}
therefore the antisymmetric component of $ F^{(31)}_{123}$ under
transposition of $12$ can always be traded by a contribution to
$T_{37}$ and one can require $F^{(31)}_{123}$ to be symmetric.

To verify that the 37 structures are necessary we have used the
following device. We consider a generic expression $F$ of the type
(\ref{eq:6.1}) with unspecified coefficient functions $F^{(i)}$. Then
$F$ is expanded in terms of {\em bare structures}, namely, structures
formed with operators $D_\mu$, as in the first equality of
(\ref{eq:6.6}). There are five such bare structures, $D_{\alpha}
D_{\alpha} D_{\beta} D_{\beta} D_{\gamma} D_{\gamma}$, $D_{\alpha}
D_{\alpha} D_{\beta} D_{\gamma} D_{\beta} D_{\gamma}$, $D_{\alpha}
D_{\alpha} D_{\beta} D_{\gamma} D_{\gamma} D_{\beta}$, $D_{\alpha}
D_{\beta} D_{\alpha} D_{\gamma} D_{\beta} D_{\gamma}$, and $D_{\alpha}
D_{\beta} D_{\gamma} D_{\alpha} D_{\beta} D_{\gamma}$ (modulo cyclic
and mirror symmetries). The reason to do this is that an expression
written in terms of bare structures is zero if and only if the
corresponding coefficient functions vanish (after imposing the
appropriate symmetry restrictions to those coefficients). That is,
there are no identities (like Jacobi or integration by parts) in terms
of bare structures, so two expressions are equal only if their
(symmetrized) coefficient functions are equal. To see that a given
structure $T_i$ is necessary, i.e., that it cannot always be written
in terms of the other structures, it is enough to expand $F$ as a
power series of $X$ in terms of bare structures and equate it to
zero. If $T_i$ were redundant, for any choice of $F^{(i)}$ there would
be choices of the other coefficient functions so that the equation
$F=0$ would hold true at each order in the series expansion. It can be
verified that this is not the case for any $T_i$ when one considers
the equations at order $X^6$.

In the six derivative case, we have not found a closed proof that the
symmetrized coefficient functions associated to an expression are
really unique. In principle there could exist non trivial identities,
that is, sets of non vanishing functions $F^{(i)}$ producing a
vanishing expression $F$. (If the $F^{(i)}$ were not symmetrized or
the $T_i$ were not all necessary, this would certainly be the case.)
To investigate this issue, we have considered large classes of
functions $F^{(i)}$ of the type encountered in $B_3$, i.e., obtained by
linear combination of functions $I_{r_1,r_2,\ldots,r_n}$, with
adjustable numerical coefficients. The corresponding expression has
been expanded in terms of bare structures and equated to zero. No
nontrivial identity has been found. Our conjecture is that the
symmetrized coefficient functions corresponding to a given expression
are unambiguous.

\begin{acknowledgments}
This work is supported in part by funds provided by the Spanish DGI
and FEDER founds with grant no. BFM2002-03218, Junta de Andaluc\'{\i}a
grant no. FQM-225, and EURIDICE with contract number
HPRN-CT-2002-00311.
\end{acknowledgments}

\appendix

\section{Summary of the notation and conventions}
\label{app:A}

In this Appendix we collect several notational conventions used in the text.
\begin{equation}
K= D_\mu^2+X \,,\quad D_\mu=\partial_\mu+V_\mu\,,\quad 
Z_{\mu\nu}:=[D_\mu,D_\nu] \,.
\end{equation}
\begin{equation}
\la~~\ra := \frac{1}{(4\pi\tau)^{d/2}} \int
d^dx\,\tr\left(~~\right) \,.
\end{equation}
Units restoration:
\begin{equation}
a_n\to \tau^n a_n\,,\qquad B_n(X)\to \tau^n B_n(\tau X) \,.
\end{equation}
Indices convention:
\begin{equation}
Y_{\mu I}= [D_\mu,Y_I] \,.
\end{equation}
Symmetric functions:
\begin{eqnarray}
I_{r_1,r_2,\ldots,r_n} &=& 
\prod_{\ell=1}^n\frac{1}{(r_\ell-1)!}
\left(\frac{\partial}{\partial X_\ell}\right)^{r_\ell-1} 
 \sum_{i=1}^ne^{X_i}\prod_{j\not= i}\frac{1}{X_i-X_j} \,.
\end{eqnarray}
\begin{eqnarray}
I_1 &=& e^X \,, \nonumber \\
I_{1,1} &=& \frac{e^{X_1}-e^{X_2}}{X_1-X_2} \,, \\
I_{1,1,1} &=& \frac{e^{X_1}-e^{X_2}}{(X_1-X_2)(X_2-X_3)}
-\frac{e^{X_1}-e^{X_3}}{(X_1-X_3)(X_2-X_3)}
 \,. \nonumber
\end{eqnarray}
Mirror transformation:
\begin{equation}
AB\to B^TA^T\,,\quad 
X_{\mu_1\ldots\mu_n} \to X_{\mu_1\ldots\mu_n} \,,\quad
Z_{\mu_1\ldots\mu_n} \to -Z_{\mu_1\ldots\mu_n} \,.
\end{equation}
Mirror symmetry convention:
\begin{equation}
Y:= \frac{1}{2}(Y+Y^T)\,.
\end{equation}


\begin{thebibliography}{10}

\bibitem{Schwinger:1951nm}
J.S. Schwinger,
\newblock Phys. Rev. 82 (1951) 664,
\newblock 

\bibitem{Gilkey:1975iq}
P.B. Gilkey,
\newblock J. Diff. Geom. 10 (1975) 601,
\newblock 

\bibitem{Atiyah:1973ad}
M. Atiyah, R. Bott and V.K. Patodi,
\newblock Invent. Math. 19 (1973) 279,
\newblock 

\bibitem{Hawking:1977ja}
S.W. Hawking,
\newblock Commun. Math. Phys. 55 (1977) 133,
\newblock 

\bibitem{Fujikawa:1980eg}
K. Fujikawa,
\newblock Phys. Rev. D21 (1980) 2848,
\newblock 

\bibitem{Ball:1989xg}
R.D. Ball,
\newblock Phys. Rept. 182 (1989) 1,
\newblock 

\bibitem{Bijnens:1996ww}
J. Bijnens,
\newblock Phys. Rept. 265 (1996) 369, hep-ph/9502335,
\newblock 

\bibitem{Bordag:2001qi}
M. Bordag, U. Mohideen and V.M. Mostepanenko,
\newblock Phys. Rept. 353 (2001) 1, quant-ph/0106045,
\newblock 

\bibitem{Callan:1994py}
J. Callan, Curtis~G. and F. Wilczek,
\newblock Phys. Lett. B333 (1994) 55, hep-th/9401072,
\newblock 

\bibitem{Bytsenko:1996bc}
A.A. Bytsenko et~al.,
\newblock Phys. Rept. 266 (1996) 1, hep-th/9505061,
\newblock 

\bibitem{Camporesi:1990wm}
R. Camporesi,
\newblock Phys. Rept. 196 (1990) 1,
\newblock 

\bibitem{Avramidi:1994zp}
I.G. Avramidi,
\newblock J. Math. Phys. 37 (1996) 374, hep-th/9406047,
\newblock 

\bibitem{Avramidi:1995ik}
I.G. Avramidi,
\newblock J. Math. Phys. 36 (1995) 5055, hep-th/9503132,
\newblock 

\bibitem{Avramidi:1997jy}
I.G. Avramidi,
\newblock Rev. Math. Phys. 11 (1999) 947, hep-th/9704166,
\newblock 

\bibitem{Dewitt:1975ys}
B.S. Dewitt,
\newblock Phys. Rept. 19 (1975) 295,
\newblock 

\bibitem{Seeley:1967ea}
R.T. Seeley,
\newblock Proc. Symp. Pure. Math. 10 (1967) 288,
\newblock 

\bibitem{Bel'kov:1996tn}
A.A. Bel'kov, A.V. Lanyov and A. Schaale,
\newblock Comput. Phys. Commun. 95 (1996) 123, hep-ph/9506237,
\newblock 

\bibitem{vandeVen:1998pf}
A.E.M. van~de Ven,
\newblock Class. Quant. Grav. 15 (1998) 2311, hep-th/9708152,
\newblock 

\bibitem{Moss:1999wq}
I.G. Moss and W. Naylor,
\newblock Class. Quant. Grav. 16 (1999) 2611, gr-qc/0101125,
\newblock 

\bibitem{Fliegner:1998rk}
D. Fliegner et~al.,
\newblock Annals Phys. 264 (1998) 51, hep-th/9707189,
\newblock 

\bibitem{Avramidi:1991je}
I.G. Avramidi,
\newblock Nucl. Phys. B355 (1991) 712,
\newblock 

\bibitem{Gusynin:1989ky}
V.P. Gusynin,
\newblock Phys. Lett. B225 (1989) 233,
\newblock 

\bibitem{Elizalde:1994bk}
E. Elizalde et~al.,
\newblock Zeta Regularization Techniques with Applications (World Scientific,
  Singapore, 1994),
\newblock 

\bibitem{Vassilevich:2003xt}
D.V. Vassilevich,
\newblock Phys. Rept. 388 (2003) 279, hep-th/0306138,
\newblock 

\bibitem{Megias:2002vr}
E. Meg{\'\i}as, E. Ruiz~Arriola and L.L. Salcedo,
\newblock Phys. Lett. B563 (2003) 173, hep-th/0212237,
\newblock 

\bibitem{Megias:2003ui}
E. Meg{\'\i}as, E. Ruiz~Arriola and L.L. Salcedo,
\newblock Phys. Rev. D69 (2004) 116003, hep-ph/0312133,
\newblock 

\bibitem{Osipov:2001bj1}
A.A. Osipov and B. Hiller,
\newblock Phys. Lett. B515 (2001) 458, hep-th/0104165,
\newblock 

\bibitem{Osipov:2001bj}
A.A. Osipov and B. Hiller,
\newblock Phys. Rev. D64 (2001) 087701, hep-th/0106226,
\newblock 

\bibitem{Salcedo:2001qp}
L.L. Salcedo,
\newblock Eur. Phys. J. direct C14 (2001) 1, arXiv:hep-th/0107133,
\newblock 

\bibitem{Vassilevich:2003yz}
D.V. Vassilevich,
\newblock Lett. Math. Phys. 67 (2004) 185, hep-th/0310144,
\newblock 

\bibitem{Barvinsky:1987uw}
A.O. Barvinsky and G.A. Vilkovisky,
\newblock Nucl. Phys. B282 (1987) 163,
\newblock 

\bibitem{Chan:1986jq}
L.H. Chan,
\newblock Phys. Rev. Lett. 57 (1986) 1199,
\newblock 

\bibitem{Salcedo:2000hp}
L.L. Salcedo,
\newblock Eur. Phys. J. C20 (2001) 147, arXiv:hep-th/0012166,
\newblock 

\bibitem{Salcedo:2000hx}
L.L. Salcedo,
\newblock Eur. Phys. J. C20 (2001) 161, arXiv:hep-th/0012174,
\newblock 

\bibitem{Gusynin:1990bu}
V.P. Gusynin and V.A. Kushnir,
\newblock Class. Quant. Grav. 8 (1991) 279,
\newblock 

\bibitem{Gusynin:1989nf}
V.P. Gusynin and V.A. Kushnir,
\newblock Sov. J. Nucl. Phys. 51 (1990) 373,
\newblock 

\bibitem{Salcedo:1998sv}
L.L. Salcedo,
\newblock Nucl. Phys. B549 (1999) 98, arXiv:hep-th/9802071,
\newblock 

\bibitem{Garcia-Recio:2000gt}
C. Garc{\'\i}a-Recio and L.L. Salcedo,
\newblock Phys. Rev. D63 (2001) 045016, arXiv:hep-th/0007183,
\newblock 

\bibitem{Caro:1993fs}
J. Caro and L.L. Salcedo,
\newblock Phys. Lett. B309 (1993) 359,
\newblock 

\bibitem{Muller:1995bu}
U. M{\"u}ller,
\newblock (1995), hep-th/9508031,
\newblock 

\bibitem{Muller:1996cq}
U. M{\"u}ller,
\newblock (1996), hep-th/9701124,
\newblock 

\bibitem{Alvarez-Gaume:2000dx}
L. \'Alvarez-Gaum\'e and S.R. Wadia,
\newblock Phys. Lett. B501 (2001) 319, hep-th/0006219,
\newblock 

\bibitem{McAvity:1991we}
D.M. McAvity and H. Osborn,
\newblock Class. Quant. Grav. 8 (1991) 603,
\newblock 

\bibitem{Salcedo:1996qy}
L.L. Salcedo and E. Ruiz~Arriola,
\newblock Ann. Phys. 250 (1996) 1, arXiv:hep-th/9412140,
\newblock 

\bibitem{Pletnev:1998yu}
N.G. Pletnev and A.T. Banin,
\newblock Phys. Rev. D60 (1999) 105017, arXiv:hep-th/9811031,
\newblock 

\end{thebibliography}

\end{document}